\RequirePackage{fix-cm}
\documentclass[smallextended]{svjour3}       
\smartqed  
\usepackage{amssymb}
\usepackage{amsmath}
\usepackage{color}
\usepackage{graphicx}
\usepackage[justification=centering]{caption}
\begin{document}

\title{Secure Quantum Private Comparison of Equality Based on Asymmetric W State
} \subtitle{}


\author{Wen-Jie Liu       \and
        Chao Liu          \and
        Hai-bin Wang       \and
        Jing-Fa Liu       \and
        Fang Wang      \and
        Xiao-Min Yuan      
}


\institute{
           W.-J. Liu   \and
           H.-B. Wang   \and
           J.-F. Liu
           \at Jiangsu Engineering Center of Network Monitoring, Nanjing University of Information Science \& Technology, Nanjing 210044, China
           \\
           \email{wenjiel@163.com}
           \and
           W.-J. Liu   \and
           C. Liu      \and
           F. Wang   \and
           X.-M. Yuan
           \at School of Computer and Software, Nanjing University of Information Science \& Technology, Nanjing 210044, China          
}

\date{Received: date / Accepted: date}

\maketitle

\begin{abstract}
Recently, Liu et al. [Opt. Commun. 284, 3160, 2011] proposed a protocol for quantum private comparison of equality (QPCE) based on symmetric $W$ state. However, Li et al. [Eur. Phys. J. D. 66, 110, 2012] pointed out that there is  a flaw of information leak, and they proposed a new protocol based on EPR pairs. While examining these two protocols, we find that there exists a same flaw: the third party (TP) can know the comparison result. In this paper, through introducing and constructing a special class of asymmetric $W$ state, a secure QPCE protocol based on this asymmetric $W$ state is presented. Analysis shows the present protocol can not only effectively avoid the information leak found by Li et al, but also ensure TP would not get any information about the comparison result.
\keywords{Quantum cryptography \and Quantum private comparison of equality \and Asymmetric $W$ state \and Information leak}
\end{abstract}

\section{Introduction}
\label{intro1}
Secure multi-party computation (SMC) is an important and fundamental branch in cryptographic field, and it deals with computing a function in a distributed network where each participant holds one of the secret inputs, and guarantees no information about one participant¡¯s secret inputs is leaked to others. In the scientific computation, comparison is a basic problem, so the private comparison is an important issue in SMC, and well-studied in classical cryptography. In 1982, Yao [1] proposed a famous protocol for the millionaires' problem, in which two millionaires wish to know who is richer without revealing the precise amount of their fortunes. Following the idea of Yao's millionaires' problem, Boudot et al. [2] subsequently proposed a protocol to decide whether two millionaires are equally rich or not. Since then, the private comparison has drawn more and more attentions [3-5]. However, the security of classical private comparison is based on the computation complexity, which is susceptible to the strong ability of quantum computation. Unlike classical private comparison, the security of quantum private comparison relies on the laws of physics, such as the Heisenberg uncertainty principle, the quantum no-cloning theorem, rather than computational assumptions. So, quantum private comparison is able to guarantee the unconditional security of the information by quantum mechanics.

The first protocol for quantum private comparison of equality (QPCE) was proposed by Yang et al. [6] in 2009, in which a one-way hash function was used to calculate the hash values of two participants' secret inputs firstly, and then these hash values were encoded into the photons of EPR pairs. In essence, its security is guaranteed by the hash function. Since then, many other QPCE protocols have been proposed, such as the QPCE protocols with non-entangled state (i.e., single photon) [7, 8], the QPCE protocols with maximally entangled state (Bell state, GHZ state, etc.) [9-13], and the QPCE protocols with non-maximally entangled state ($W$ state, cluster state, $\chi$-type state, etc.) [14-18]. Recently, Liu et al. [18] proposed an efficient QPCE protocol with three-particle symmetric $W$ state $\frac{1}{\sqrt{3}}(|100\rangle+|010\rangle+|001\rangle)$ (referred to as the LWJ11 protocol hereafter). To our knowledge, this is the only one that utilized $W$ state as quantum resource in the QPCE protocol. Unfortunately, in 2012, Li et al. [19] pointed out that a flaw of information leak is existent in the LWJ11 protocol: one participant can estimate the other's private bit successfully with probability $2/3$, rather than $1/2$. Subsequently, they presented a new protocol with EPR pairs (referred to as the LWG12 protocol hereafter). Thus, strictly speaking, there has been no secure QPCE protocol based on $W$ state to be published by far. On the other hand, we find that there exists a same flaw in the LWJ11 and LWG12 protocols, i.e., the third party (TP) can know the comparison result, which is contrary to what they claimed: ``TP cannot know the comparison result".

In this paper, we aim to propose a secure QPCE protocol based on $W$ state to avoid all of these flaws discussed above. Firstly, we introduce a special class of asymmetric $W$ state, and give its construction framework from the perspective of quantum reversible logic circuits. And then, by utilizing the $W$ states, a new secure QPCE protocol is presented. The new protocol not only holds the same efficiency as the LWJ11 protocol, but also avoids the information leak pointed out in Ref. [19]. What is more, the present protocol can ensure TP cannot know any information about the comparison result, which avoids the flaw existent in the LWJ11 and LWG12 protocols.

The structure of this paper is organized as follows. In Section 2, we briefly review and analyze the LWJ11 and LWG12 protocols. In Section 3, the asymmetric $W$ state is introduced, and a new secure QPCE protocol is proposed. And the security of the present protocol is analyzed in Section 3, and a brief discussion and conclusion is given finally in Section 4.

\section{Review of the LWJ11 and LWG12 protocols and their flaw}
\label{sec:2}
\subsection{Review of the LWJ11 and LWG12 protocols}
In Ref. [18], Liu et al. proposed a QPCE protocol based on symmetric $W$ state. In the protocol, Alice and Bob are supposed as two participants, who have the secret inputs $X=\sum^{N-1}_{i=0}x_{i}2^{i}$, $Y=\sum^{N-1}_{i=0}y_{i}2^{i}$, respectively, where $x_{i},y_{i}\in\{0,1\}$, $2^{N-1}\leq max\{X,Y\}\leq2^{N}$. Obviously, the binary representations of $X$ and $Y$ are $(x_{0},x_{1},...,x_{N-1})$, $(y_{0},y_{1},...,y_{N-1})$, respectively. And the main procedures of the LWJ11 protocol are as follows,

\textbf{Step 1}. Alice and TP (Bob and TP), use a QKD protocol to establish a common secret key $K_{AC}$ $(K_{BC})$. Alice and Bob agree that $|01\rangle, |10\rangle,|1\rangle$ $(|00\rangle,|0\rangle)$ represent information `1' (`0').

\textbf{Step 2.} Alice (Bob) prepares $N$ symmetric $W$ states, each of which is randomly chosen from $|\phi_{1}\rangle=\frac{1}{\sqrt{3}}(|100\rangle+|010\rangle+|001\rangle)_{123}$, $|\phi_{2}\rangle=\frac{1}{\sqrt{3}}(|10+\rangle+|01+\rangle+|00-\rangle)_{123}$. Then Alice (Bob) performs $i\sigma_{y}=|0\rangle\langle1|-|1\rangle\langle0|$ operation on the third photon of $i$th $W$ state when $x_{i}=1$ $(y_{i}=1)$, where $i=0,1,\ldots,N-1$. After that, Alice (Bob) prepares $N^{'}$ $W$ states from one of the two states $|\phi_{1}\rangle$, $|\phi_{2}\rangle$ randomly, and inserts them into her (his) $N$ $W$ states at the same positions. Then Alice and Bob exchange the third photons of all $W$ states and remain the other photons on hand. Subsequently, Alice (Bob) announces publicly the $N+N^{'}$ initial states prepared by her (him). If the initial state is $|\phi_{2}\rangle$, Bob (Alice) performs a Hadamard operation $(H)$ on the third photons arrived from Alice (Bob).
\begin{equation}
H=\frac{1}{\sqrt{2}}(|0\rangle\langle0|+|1\rangle\langle0|+|0\rangle\langle1|-|1\rangle\langle1|),
\end{equation}
Otherwise, he (she) does nothing. Finally, Alice and Bob take out those $2N^{'}$ $W$ states and perform the eavesdropping checking, respectively. The detailed procedures are omitted here.

\textbf{Step 3}. If there is no eavesdropper, Alice (Bob) makes single-particle measurement in the $Z$-basis$\{|0\rangle,|1\rangle\}$ on every photon on her (his) hand. The measurement outcome about the first and second photons of every $W$ state prepared by Alice (Bob) is denoted by $M^{A1}_{i}$ $(M^{B1}_{i})$. If $M^{A1}_{i}$ $(M^{B1}_{i})$ is $|01\rangle$ or $|10\rangle$, then $C^{A1}_{i}(C^{B1}_{i})=1$; if $M^{A1}_{i}$ $(M^{B1}_{i})$ is $|00\rangle$, then $C^{A1}_{i}(C^{B1}_{i})=0$. And the outcome about the third photons of every $W$ states prepared by Alice (Bob) is represented as $M^{A2}_{i}$ $(M^{B2}_{i})$. When $M^{A2}_{i}$ $(M^{B2}_{i})$ is $|0\rangle$, then $C^{A2}_{i}(C^{B2}_{i})=0$; When $M^{A2}_{i}$ $(M^{B2}_{i})$ is $|1\rangle$, then $C^{A2}_{i}(C^{B2}_{i})=1$. Finally, Alice (Bob) calculates $C^{A}_{i}=C^{A1}_{i}\oplus C^{B2}_{i}$, $(C^{B}_{i}=C^{B1}_{i}\oplus C^{A2}_{i})$, and gets $C_{A}=(C^{A}_{0},C^{A}_{1},\ldots ,C^{A}_{N-1})$ $(C_{B}=(C^{B}_{0},C^{B}_{1},\ldots ,C^{B}_{N-1}))$. Here $i=0,1,\cdots N-1$.

\textbf{Step 4}. Alice (Bob) prepares an $L$-length random sequence $C_{A^{'}}=(C^{A^{'}}_{0},C^{A^{'}}_{1}\\,\ldots ,C^{A^{'}}_{L-1})$ $(C_{B^{'}}=(C^{B^{'}}_{0},C^{B^{'}}_{1},\ldots ,C^{B^{'}}_{L-1}))$, where $C^{A^{'}}_{i},C^{B^{'}}_{i}\in \{0,1\}$, $i=0,1,\ldots ,L-1$, and sends $C_{A^{'}}$ $(C_{B^{'}})$ to Bob (Alice). Alice inserts $C_{A^{'}}$ into $C_{A}$ to get $C_{A^{''}}$ and sends the inserted positions sequence $S_{q}$ to Bob. Bob inserts $C_{B^{'}}$ into $C_{B}$ in terms of $S_{q}$ to get $C_{B^{''}}$. Then Alice and Bob use $K_{AC}$, $K_{BC}$ to encrypt $C_{A^{''}}$, $C_{B^{''}}$, get $E(C_{A^{''}})$, $E(C_{B^{''}})$, respectively, and send both of them to TP.

\textbf{Step 5}. After using $K_{AC}$, $K_{BC}$ to decrypt $E(C_{A^{''}})$, $E(C_{B^{''}})$ and getting $C_{A^{''}}$, $C_{B^{''}}$, TP calculates $R^{'}=\sum^{N-1}_{i=0}(C^{A}_{i}\oplus C^{B}_{i})+\sum^{L-1}_{i=0}(C^{A^{'}}_{i}\oplus C^{B^{'}}_{i})$, and he sends $R^{'}$ to Alice and Bob.

\textbf{Step 6}. After receiving $C_{A^{'}}$, $C_{B^{'}}$, $R^{'}$, Alice and Bob calculate $R=R^{'}-\sum^{L-1}_{i=0}(C^{A^{'}}_{i}\oplus C^{B^{'}}_{i})$, respectively. If $R=0$, they know $X=Y$; otherwise, $X\neq Y$.

In Ref. [19], Li et al. pointed out that a flaw of information leak is existent in the LWJ11 protocol, that is, a participant can estimate the other's every private information correctly with probability $2/3$. Then they proposed a new protocol using EPR pairs instead of the original symmetric $W$ states. The whole protocol consists of six steps, whose main procedures are similar as those of LWJ11 protocol. The major modification is as follows,
\begin{description}
  \item[(i)] In the whole protocol, the EPR pairs $|\varphi^{+}\rangle=\frac{1}{\sqrt{2}}(|01\rangle+|10\rangle)$ are utilized as information carrier instead of the original $W$ states.
  \item[(ii)] In order to save quantum bits, decoy photons, rather than entangled $W$ states, are utilized to perform eavesdropping checking.
\end{description}

\subsection{The flaw}
In order to prevent TP from getting the comparison result, a mix-up solution is used in both of the LWJ11 and LWG12 protocols. More specifically, as depicted in Step 4, Alice (Bob) prepares the random sequence $C_{A^{'}}$ $(C_{B^{'}})$, and sends it to Bob (Alice). Then Alice inserts $C_{A^{'}}$ into $C_{A}$, and gets $C_{A^{''}}$. At the same time, Bob inserts $C_{B^{'}}$ into $C_{B}$ at the same positions in terms of Alice's notification, and gets $C_{B^{''}}$. For simplicity, we denote $C_{A^{''}}$, $C_{B^{''}}$ as below, $C_{A^{''}}=C_{A}\|C_{A^{'}}$, $C_{B^{''}}=C_{B}\|C_{B^{'}}$ (`$\|$' denotes the collection of two sequences). Subsequently, $C_{A^{''}}$ and $C_{B^{''}}$ will be sent to TP by utilizing QKD method in Step 4. And then, TP calculates the bit-wise exclusive-OR operation between $C_{A^{''}}$ and $C_{B^{''}}$, and gets $R^{'}=\sum^{N-1}_{i=0}(C^{A}_{i}\oplus C^{B}_{i})+\sum^{L-1}_{i=0}(C^{A^{'}}_{i}\oplus C^{B^{'}}_{i})$. On the surface, TP only know the result of $R^{'}$ rather than $R=\sum^{N-1}_{i=0}(C^{A}_{i}\oplus C^{B}_{i})$, so he cannot know the comparison result.

 But unfortunately, that is just not the case. As we all know, the classical information transmitted through public channel can be arbitrarily obtained without being detected. Since $C_{A^{'}}$, $C_{B^{'}}$ and $S_{q}$ are sent through the classical channel in Step 4, that means TP can obtain them without be detected. There are two ways for TP to gain the comparison result. (1) TP calculates $R^{''}=\sum^{L-1}_{i=0}(C^{A^{'}}_{i}\oplus C^{B^{'}}_{i})$ according to $C_{A^{'}}$ and $C_{B^{'}}$, and gets the result by calculating $R=R^{'}-R^{''}$. (2) TP gets rid of the mix-up sequences $C_{A^{'}}$, $C_{B^{'}}$ from $C_{A^{''}}$, $C_{B^{''}}$, respectively, and obtains $C_{A}$, $C_{B}$ according to the position sequence $S_{q}$, and then he calculates $R=\sum^{N-1}_{i=0}(C^{A}_{i}\oplus C^{B}_{i})$ to get the comparison result. In a word, TP can gain the comparison result, which is contrary to what they claimed: ``TP cannot know the comparison result".

\section{The proposed protocol based on asymmetric $W$ state}
\label{sec:3}
\subsection{The asymmetric $W$ state and its quantum circuits construction}
Unlike the maximally entangled state, the $W$ state has some nice entanglement properties, it is maximally robust under disposal of any one of the three qubits. The class of states $|W_{n}\rangle$ is such one [20],
\begin{equation}
|W_{n}\rangle=\frac{1}{\sqrt{2+2n}}(|100\rangle+\sqrt{n}e^{i\gamma}|010\rangle+\sqrt{n+1}e^{i\delta}|001\rangle)_{123},
\end{equation}
where $n$ is a real number, $\gamma$ and $\delta$ are phases. As we shall see, the asymmetric states can be used for perfect superdense coding in our protocol. In particular, if we take $n=1$ for simplicity and set phases to zero (i.e., $\gamma=0$, $\delta=0$), then the $|W_{1}\rangle$ state can be gotten,
\begin{equation}
|W_{1}\rangle=\frac{1}{2}(|100\rangle+|010\rangle+\sqrt{2}|001\rangle)_{123}.
\end{equation}
From the perspective of quantum reversible logic circuits [21], we design a framework of quantum circuit by using NOT gate, controlled-NOT (CNOT) gate and Hadamard gate to construct the asymmetric $W$ state $|W_{1}\rangle$. The detailed circuit representation is shown in Fig. 1.

\begin{figure}
  \centering
  \includegraphics[width=8.5cm]{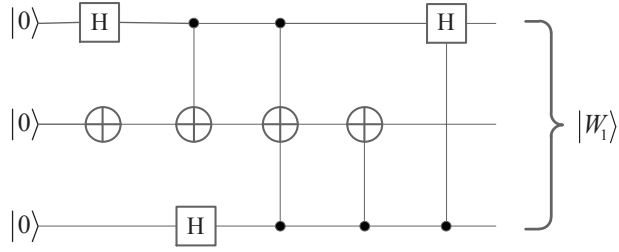}\\
  \caption{The quantum circuit framework generating the $|W_{1}\rangle$ state.}\label{Fig. 1}
\end{figure}

\subsection{The QPCE protocol based on asymmetric $W$ state}
In order to avoid the flaw of information leak which Li et al. pointed out, as well as prevent TP from getting the comparison result, we present a secure QPCE protocol based on asymmetric $W$ state, and the detailed procedures are as below.

\textbf{Prerequisite}. Alice and Bob use a QKD protocol to establish two common secret keys $K_{AB}$; Alice and TP (Bob and TP) get $K_{AT}$ $(K_{BT})$ in the same way. Alice and Bob agree that $|01\rangle, |10\rangle,|1\rangle$  represent the information `1'; $|00\rangle,|0\rangle$ represent the information `0'.

\textbf{Step 1}. Alice and Bob prepare an ordered $N$ three-qubit asymmetric $W$ states sequence, respectively, which are called $S_{A}$, $S_{B}$,
\begin{equation}
\left\{
\begin{array}{rcl}
S_{A}=[P^{A_{1}}_{1},P^{A_{2}}_{1},P^{A_{3}}_{1},P^{A_{1}}_{2},P^{A_{2}}_{2},P^{A_{3}}_{2},\ldots,P^{A_{1}}_{N},P^{A_{2}}_{N},P^{A_{3}}_{N}]~\\
S_{B}=[P^{B_{1}}_{1},P^{B_{2}}_{1},P^{B_{3}}_{1},P^{B_{1}}_{2},P^{B_{2}}_{2},P^{B_{3}}_{2},\ldots,P^{B_{1}}_{N},P^{B_{2}}_{N},P^{B_{3}}_{N}].
\end{array}
\right.
\end{equation}
Here, the subscripts $1,2,...,N$ indicate the orders of $W$ states in the sequence and superscripts $A_{1},A_{2},A_{3} (B_{1},B_{2},B_{3})$ represent three different particles in one $W$ state. And each $W$ state is chosen from $\{|W_{1}\rangle,|W^{'}_{1}\rangle\}$ randomly.
\begin{equation}
|W_{1}\rangle=\frac{1}{2}(|100\rangle+|010\rangle+\sqrt{2}|001\rangle)_{123},
\end{equation}
\begin{equation}
|W^{'}_{1}\rangle=\frac{1}{2}(|10+\rangle+|01+\rangle+\sqrt{2}|00-\rangle)_{123}.
\end{equation}
Then Alice (Bob) perform unitary operation $I$ or $\sigma_{x}$ on the third photon of $i$th $W$ state in terms of her (his) secret inputs $(x_{0},x_{1},...,x_{N-1})$ $((y_{0},y_{1},...,y_{N-1}))$. If $x_{i}=1(y_{i}=1)$, Alice (Bob) performs $\sigma_{x}=|0\rangle\langle1|+|1\rangle\langle0|$ operation; otherwise, she (he) performs $I=|0\rangle\langle0|+|1\rangle\langle1|$ operation, where $i=0,1,\ldots,N-1$. After that, Alice (Bob) takes particles 3 from each state in $S_{A}$ $(S_{B})$ to form an ordered particle sequence $S^{A}_{3}$ $(S^{B}_{3})$
\begin{equation}
\left\{
\begin{array}{rcl}
S^{A}_{3}=[P^{A_{3}}_{1},P^{A_{3}}_{2},\ldots,P^{A_{3}}_{N}]~\\
S^{B}_{3}=[P^{B_{3}}_{1},P^{B_{3}}_{2},\ldots,P^{B_{3}}_{N}].
\end{array} \right.
\end{equation}
For preventing eavesdropping, Alice (Bob) prepares a set of decoy photons $D_{A}$ $(D_{B})$ randomly in four nonorthogonal photon states $\{|0\rangle$,$|1\rangle$,$|+\rangle$,$|-\rangle\}$, and inserts $D_{A}$ $(D_{B})$ into the third photon sequence $S^{A}_{3}$ $(S^{B}_{3})$ at random positions to form a new sequences $S^{A*}_{3}$ $(S^{B*}_{3})$. Then Alice and Bob exchange $S^{A*}_{3}$ and $S^{B*}_{3}$.

\textbf{Step 2}. After receiving $S^{B*}_{3}$ $(S^{A*}_{3})$, Alice (Bob) announces publicly her (his) $N$ initial asymmetric $W$ states. If the initial state of Bob's (Alice's) is $|W^{'}_{1}\rangle$, Alice (Bob) will perform a Hadamard operation (see Eq. (1)) on the third photon on her (his) hand; if the state is $|W_{1}\rangle$, they do nothing. To guarantee the security of the quantum channel, Alice (Bob) will perform the eavesdropping check. The checking procedure of Alice-Bob quantum channel is: (i) Alice informs Bob the positions and the measurement bases of the decoy photons. (ii) Bob performs single-photon measurement and publishes his measurement outcomes. (iii) Alice analyzes the error rate, if the error rate is higher than the threshold they preset, she aborts the protocol and restarts from the preparing step; otherwise, the quantum channel of Alice-Bob is secure. On the same time, Bob utilizes the same method to check security of Bob-Alice quantum channel.

\textbf{Step 3}. Alice (Bob) discards the decoy photons in $S^{A*}_{3}$ $(S^{B*}_{3})$ and makes the $Z$-basis measurement on every photon of her (his) hand. The measurement outcome of Alice's (Bob's) first and second photons in the $i$th $W$ state is denoted as $M^{A1}_{i}$ $(M^{B1}_{i})$, then the value of $C^{A1}_{i}$ $(C^{B1}_{i})$ can be computed as follows,
\begin{equation}
C^{A1}_{i}(C^{B1}_{i})=\left\{
\begin{array}{rcl}
0,~M^{A1}_{i}(M^{B1}_{i})=|00\rangle~~~~~~~~~~~\\
1,~M^{A1}_{i}(M^{B1}_{i})=|01\rangle~or~|10\rangle.
\end{array} \right.
\end{equation}
And $C^{B2}_{i}$($C^{A2}_{i}$) can be gotten according to Alice's (Bob's) outcome of the third photon $M^{B2}_{i}$ ($M^{A2}_{i}$),
\begin{equation}
C^{B2}_{i}(C^{A2}_{i})=\left\{
\begin{array}{rcl}
0,~M^{B2}_{i}(M^{A2}_{i})=|0\rangle~\\
1,~M^{B2}_{i}(M^{A2}_{i})=|1\rangle.
\end{array} \right.
\end{equation}
Then, Alice (Bob) continue to calculate $C^{A}_{i}=C^{A1}_{i}\oplus C^{B2}_{i}$ $(C^{B}_{i}=C^{B1}_{i}\oplus C^{A2}_{i})$, and denotes $C_{A}=(C^{A}_{0},C^{A}_{1},\ldots ,C^{A}_{N-1})$ $(C_{B}=(C^{B}_{0},C^{B}_{1},\ldots ,C^{B}_{N-1}))$.

\textbf{Step 4}. Alice (Bob) prepares an $L$-length random binary sequence $C^{'}_{A}=(C^{A^{'}}_{0},C^{A^{'}}_{1},\ldots ,C^{A^{'}}_{N-1})$ $(C^{'}_{B}=(C^{B^{'}}_{0},C^{B^{'}}_{1},\ldots ,C^{B^{'}}_{N-1}))$, where $C^{A^{'}}_{i},C^{B^{'}}_{i}\in \{0,1\}$, $i=0,1,\ldots ,L-1$, After that, Alice (Bob) uses $K_{AB}$ to encrypt $C^{'}_{A}$ $(C^{'}_{B})$, and sends the result $E(C^{'}_{A})$ $(E(C^{'}_{B}))$ to Bob (Alice). Having received $E(C^{'}_{B})$ $(E(C^{'}_{A}))$, Alice (Bob) uses $K_{AB}$ to decrypt $E(C^{'}_{B})$ $(E(C^{'}_{A}))$ and gets $C^{'}_{B}$ $(C^{'}_{A})$. Then Alice inserts $C^{'}_{A}$ into $C_{A}$ at random positions to obtain a new sequence named $C^{''}_{A}$, she also records the inserted positions sequence $S_{q}$. Then Alice uses $K_{AB}$ to encrypt $S_{q}$ and sends the result $E(S_{q})$ to Bob. Bob decrypt $E(S_{q})$ to get $S_{q}$, and inserts $C^{'}_{B}$ into $C_{B}$ in terms of $S_{q}$ for getting $C^{''}_{B}$. Subsequently, Alice and Bob use $K_{AT}$, $K_{BT}$ to encrypt $C^{''}_{A}$, $C^{''}_{B}$, get $E(C^{''}_{A})$, $E(C^{''}_{B})$, and send them to TP, respectively.

\textbf{Step 5}. TP decrypts $E(C^{''}_{A})$, $E(C^{''}_{B})$ by using $K_{AT}$, $K_{BT}$, and gets $C^{''}_{A}$, $C^{''}_{B}$. Then he calculates $R^{'}=\sum^{N-1}_{i=0}(C^{A}_{i}\oplus C^{B}_{i})+\sum^{L-1}_{i=0}(C^{A^{'}}_{i}\oplus C^{B^{'}}_{i})$, and sends $R^{'}$ to Alice and Bob, respectively.

\textbf{Step 6}. After receiving $C^{'}_{A}$, $C^{'}_{B}$, $R^{'}$, Alice and Bob calculate $R=R^{'}-\sum^{L-1}_{i=0}(C^{A^{'}}_{i}\oplus C^{B^{'}}_{i})$, respectively. If $R=0$, Alice and Bob will know $X=Y$; otherwise, they get $X\neq Y$.

An example is given for better understanding the presented protocol. Suppose that the $i$th secret inputs of binary representations of $X$ and $Y$ are `0' and `0'. The asymmetric $W$ state of Alice (Bob) prepared is $|W_{1}\rangle$ $(|W^{'}_{1}\rangle)$. After performing unitary operations and the Hadamard operation, $|W_{1}\rangle$ and $|W^{'}_{1}\rangle$ evolve to a same state $|W^{1}\rangle_{0}=\frac{1}{2}(|100\rangle+|010\rangle+\sqrt{2}|001\rangle)$. so $C^{A1}_{i}=1$, $C^{B2}_{i}=0$, $C^{B1}_{i}=1$, $C^{A2}_{i}=0$ can be calculated according to the single-particle measurement outcomes. Thus, $C^{A}_{i}=C^{A1}_{i}\oplus C^{B2}_{i}=1$, $C^{B}_{i}=C^{B1}_{i}\oplus C^{A2}_{i}=1$, $R_{i}=C^{A}_{i}\oplus C^{B}_{i}=1\oplus 1=0$. According to $R_{i}$, Alice and Bob know that $x_{i}=y_{i}$. For $i=0$ to $N-1$, Alice and Bob use same method to compare whether $x_{i}$, $y_{i}$ are equal or not. In Tabel 1, we give all different cases of $x_{i},y_{i}$'s values in the proposed QPCE protocol.

\begin{table}[h]\scriptsize
\caption{All different cases of $x_{i}$, $y_{i}$'s values}
\label{tab:1}       
\begin{tabular}{lllllllllllll}
\hline\noalign{\smallskip}
$x_{i}$ & $y_{i}$ & $M^{A1}_{i}$ & $M^{B2}_{i}$ & $M^{B1}_{i}$ & $M^{A2}_{i}$ & $C^{A1}_{i}$ & $C^{B2}_{i}$ & $C^{B1}_{i}$ & $C^{A2}_{i}$ & $C^{A}_{i}$ & $C^{B}_{i}$ & $C_{i}$\\
\noalign{\smallskip}\hline\noalign{\smallskip}
0 & 0 & $|01\rangle or |10\rangle$ & $|0\rangle$ & $|01\rangle or |10\rangle$ & $|0\rangle$ & 1 & 0 & 1 & 0 & 1 & 1 & 0\\
 &  & $|01\rangle or |10\rangle$ & $|1\rangle$ & $|00\rangle$ & $|0\rangle$ & 1 & 1 & 0 & 0 & 0 & 0 & 0\\
 &  & $|00\rangle$ & $|0\rangle$ & $|01\rangle or |10\rangle$ & $|1\rangle$ & 0 & 0 & 1 & 1 & 0 & 0 & 0\\
 &  & $|00\rangle$ & $|1\rangle$ & $|00\rangle$ & $|1\rangle$ & 0 & 1 & 0 & 1 & 1 & 1 & 0\\
0 & 1 & $|01\rangle or |10\rangle$ & $|1\rangle$ & $|01\rangle or |10\rangle$ & $|0\rangle$ & 1 & 1 & 1 & 0 & 0 & 1 & 1\\
 &  & $|01\rangle or |10\rangle$ & $|0\rangle$ & $|00\rangle$ & $|0\rangle$ & 1 & 0 & 0 & 0 & 1 & 0 & 1\\
 &  & $|00\rangle$ & $|0\rangle$ & $|00\rangle$ & $|1\rangle$ & 0 & 0 & 0 & 1 & 0 & 1 & 1\\
 &  & $|00\rangle$ & $|1\rangle$ & $|01\rangle or |10\rangle$ & $|1\rangle$ & 0 & 1 & 1 & 1 & 1 & 0 & 1\\
1 & 0 & $|01\rangle or |10\rangle$ & $|0\rangle$ & $|01\rangle or |10\rangle$ & $|1\rangle$ & 1 & 0 & 1 & 1 & 1 & 0 & 1\\
 &  & $|00\rangle$ & $|0\rangle$ & $|01\rangle or |10\rangle$ & $|0\rangle$ & 0 & 0 & 1 & 0 & 0 & 1 & 1\\
 &  & $|00\rangle$ & $|1\rangle$ & $|00\rangle$ & $|0\rangle$ & 0 & 1 & 0 & 0 & 1 & 0 & 1\\
 &  & $|01\rangle or |10\rangle$ & $|1\rangle$ & $|00\rangle$ & $|1\rangle$ & 1 & 1 & 0 & 1 & 0 & 1 & 1\\
 1 & 1 & $|01\rangle or |10\rangle$ & $|1\rangle$ & $|01\rangle or |10\rangle$ & $|1\rangle$ & 1 & 1 & 1 & 1 & 0 & 0 & 0\\
 &  & $|01\rangle or |10\rangle$ & $|0\rangle$ & $|00\rangle$ & $|1\rangle$ & 1 & 0 & 0 & 1 & 1 & 1 & 0\\
 &  & $|00\rangle$ & $|1\rangle$ & $|01\rangle or |10\rangle$ & $|0\rangle$ & 0 & 1 & 1 & 0 & 1 & 1 & 0\\
 &  & $|00\rangle$ & $|0\rangle$ & $|00\rangle$ & $|0\rangle$ & 0 & 0 & 0 & 0 & 0 & 0 & 0\\
\noalign{\smallskip}\hline
\end{tabular}
\end{table}

\section{Security Analysis}
\label{sec 4.}
The proposed protocol holds the same security as the LWJ11 and LWG12 protocols, and it is secure against the well-known attacks (e.g., intercept-resend attack, measurement-resend attack, and entanglement-measure attack, etc.). Moreover, the protocol can avoid the flaw of information leak pointed out by Li et al., and ensure TP cannot get comparison result. Here, we just have a brief analysis on these two aspects.

\begin{flushleft}
\textbf{(1) The information leak Li et al. pointed out can be avoided}
\end{flushleft}

Without loss of generality, we suppose that Bob is a dishonest participant who attempts to obtain Alice secret inputs. The only way for Bob is to use the third particles of Alice's asymmetric $W$ states (i.e., $S^{A}_{3}$) sent to him. The detailed description is as follows,

In Step 1, Alice prepares $|W_{1}\rangle=\frac{1}{2}(|100\rangle+|010\rangle+\sqrt{2}|001\rangle)_{123}$,
$|W^{'}_{1}\rangle=\frac{1}{2}(|10+\rangle+|01+\rangle+\sqrt{2}|00-\rangle)_{123}$ randomly. For simplicity, we take the first state $|W_{1}\rangle$ as an example. After Alice's encoding operation on the third particle, $|W_{1}\rangle$ evolves $|W^{1}\rangle_{0}$ or $|W^{1}\rangle_{1}$ in terms of Alice's secret inputs.
\begin{equation}
\left\{
\begin{array}{rcl}
|W^{1}\rangle_{0}=\frac{1}{2}(|100\rangle+|010\rangle+\sqrt{2}|001\rangle)_{123}~~~~~x_{i}=0\\
|W^{1}\rangle_{1}=\frac{1}{2}(|101\rangle+|011\rangle+\sqrt{2}|000\rangle)_{123}~~~~~x_{i}=1.
\end{array} \right.
\end{equation}
As specified in Step 3, Bob will perform a single-particle measurement on the third particle. As shown in Eq. (10), if Bob can deduce the final state is $|W^{1}\rangle_{0}$ or $|W^{1}\rangle_{1}$, he will know the Alice secret input ($x_{i}=0$ or $x_{i}=1$). At the same time, the reduced density operator of Bob's photon is
\begin{equation}
\begin{aligned}
\rho^{B}_{0}&=tr_{12}(\frac{|100\rangle+|010\rangle+\sqrt{2}|001\rangle}{2}\otimes\frac{\langle100|+\langle010|+\sqrt{2}\langle001|}{2})\\
&=\frac{1}{2}|0\rangle\langle0|+\frac{1}{2}|1\rangle\langle1|
=\begin{bmatrix}\frac{1}{2} &0\\0 &\frac{1}{2}\end{bmatrix}
~~~x_{i}=0\\
\end{aligned}
\end{equation}

\begin{equation}
\begin{aligned}
\rho^{B}_{1}&=tr_{12}(\frac{|101\rangle+|011\rangle+\sqrt{2}|000\rangle}{2}\otimes\frac{\langle101|+\langle011|+\sqrt{2}\langle000|}{2})\\
&=\frac{1}{2}|0\rangle\langle0|+\frac{1}{2}|1\rangle\langle1|
=\begin{bmatrix}\frac{1}{2} &0\\0 &\frac{1}{2}\end{bmatrix}
~~~x_{i}=1
\end{aligned}
\end{equation}

To estimate Alice's $i$th private bit, Bob must estimate whether the system in his possession is described by $\rho^{B}_{0}$ or $\rho^{B}_{1}$. Using the optimal measurement [22, 23], the maximum probability that Bob estimates Alice's $i$th private bit correctly is
\begin{equation}
P^{max}=\frac{1}{2}+\frac{1}{4}Tr|\rho^{B}_{0}-\rho^{B}_{1}|=\frac{1}{2}
\end{equation}
where
$|\rho^{B}_{0}-\rho^{B}_{1}|=\sqrt{(\rho^{B}_{0}-\rho^{B}_{1})^{\dag}(\rho^{B}_{0}-\rho^{B}_{1})}$,
and $(\rho^{B}_{0}-\rho^{B}_{1})^{\dag}$ is the Hermitian conjugate
of the $(\rho^{B}_{0}-\rho^{B}_{1})$ matrix. That is, Bob can estimate Alice's private bit successfully with $1/2$ probability. For instance, if $M^{A2}_{i}$ is $|0\rangle$, Bob would guess $x_{i}=0$ with the
$1/2$ probability; if $M^{A2}_{i}$ is $|1\rangle$, he
would guess $x_{i}=1$ with the $1/2$ probability, too. That means, Bob cannot get Alice's secret inputs by measuring the third particles $S^{A}_{3}$. So, in our protocol, there is no flaw of information leak presented in Ref.[19]

\begin{flushleft}
\textbf{(2) TP cannot know the comparison result}
\end{flushleft}

As analyzed in Section 2.2, the only chance for TP to know the comparison result is to get the mix-up sequences $C^{'}_{A}$, $C^{'}_{B}$ or the inserted positions sequence $S_{q}$. In our scenario, these sequences $C^{'}_{A}$, $C^{'}_{B}$ and $S_{q}$ are encrypted by the QKD key $(K_{AB})$, which is only shared with Alice and Bob. And the QKD protocol has already been proven to be unconditionally secure [24, 25], that means, none can get any information about $C^{'}_{A}$, $C^{'}_{B}$ and $S_{q}$.

Let us check the feasibility of two possible ways for TP to gain the comparison result. (1) TP tries to intercept $C^{'}_{A}$ and $C^{'}_{B}$ from the public channel between Alice and Bob, then calculates $R^{''}=\sum^{L-1}_{i=0}(C^{A^{'}}_{i}\oplus C^{B^{'}}_{i})$, and finally gets the result by calculating $R=R^{'}-R^{''}$. However, in our new protocol, $C^{'}_{A}$ and $C^{'}_{B}$ are encrypted into $E(C^{'}_{A})$, $E(C^{'}_{B})$ by using the QKD key $(K_{AB})$ before they are sent to the counterpart through the public channel. So, TP cannot get $C^{'}_{A}$ and $C^{'}_{B}$ from $E(C^{'}_{A})$ and $E(C^{'}_{B})$, and then he cannot calculate the result of $R^{''}$, that means TP will not get the comparison result $R$. (2) TP attempts to get rid of the mix-up sequence $C^{'}_{A}$ and $C^{'}_{B}$ according to the position sequence $S_{q}$, gets $C_{A}$ and $C_{B}$, and then calculates $R=\sum^{N-1}_{i=0}(C^{A}_{i}\oplus C^{B}_{i})$. This attack strategy does not work either. As described in Step 4 of our protocol, ``Alice uses $(K_{AB})$ to encrypt $S_{q}$ and sends the result $E(S_{q})$ to Bob". That means TP cannot get $S_{q}$, and undoubtedly he cannot obtain the result $R$. In summary, our protocol can prevent TP from knowing the comparison result.

\section{Discussion and Conclusion}
\label{sec 5.}
Up to now, many QPCE protocols based on maximally entangled states (e.g., Bell state, GHZ state, GHZ-like state) have been proposed. Comparing with the maximally entangled states, the non-maximally entangled states (e.g., $W$ state, cluster state, $\chi$-type state) have the stronger non-classicality, and they are more robust against particle losses than the corresponding maximally entangled states. Moreover, since we can directly use these non-maximally entangled states and do not need to distill maximally entangled states from them, it is more convenient and economical in the practical experiment. From the perspective of the robustness and the physical implementation, using the non-maximally entangled state as the quantum resource to design QPCE protocol may be a best choice.

To guarantee the security of quantum channel, the eavesdropping checking is indispensable. As we all know, there are two common methods, the first is to utilize the correlation of entangled states, and the second is to use the decoy photons. In the LWJ11 protocol, they utilize the $W$ states to perform eavesdropping checking, while decoy photons are used to replace $W$ states for eavesdropping checking in the LWG12 protocol. From the view of qubit efficiency, using decoy photons to perform eavesdropping checking is obviously better than using entangled states. Similar to the LWG12 protocol, we choose the decoy photons to check the channel security. On the other hand, many QPCE protocols, including the LWJ11 and LWG12 protocols, adopt the QKD method to enhance the security of quantum private comparison. Undoubtedly, this will cost more quantum resource, but it is the price we have to pay for the higher safety requirements.

In this paper, based on the asymmetric $W$ state $|W_{1}\rangle$ and the QKD encryption method, a new secure QPCE protocol is proposed. Compared with other analogous protocols, it not only holds the same security as that in Ref. [18], but also can avoid the information leak pointed out in Ref. [19]. In addition, the protocol can ensure that TP cannot know the comparison result. On the other hand, the protocol inherits the efficiency of $W$ state, which is discussed in Ref. [18]. What is more, TP is only to do some exclusive-OR operations, the two participants are only required to perform the simpler single-particle measurement, which is more economical and feasible to be implemented within present technologies. Our two-party protocol is a simple QPCE model, which can be generalized to the case of multi-party and be extended to solve the Yao's millionaires' problem.

\begin{acknowledgements}
This work is supported by the National Nature Science Foundation of China (Grant Nos. 61103235, 61170321, 61373016 and 61373131), the Priority Academic Program Development of Jiangsu Higher Education Institutions (PAPD), the Natural Science Foundation of Jiangsu Province, China (BK2010570), and the Practice Inovation Trainng Program Projects for the Jiangsu College Students (201310300018Z).
\end{acknowledgements}



\end{document}